# Building A Smart Academic Advising System Using Association Rule Mining

Raed Shatnawi
Jordan University of Science and Technology
+962795285056
raedamin@just.edu.jo

Qutaibah Althebyan
Jordan University of Science and Technology
+962796536277
qaalthebyan@just.edu.jo

Baraq Ghalib & Mohammed Al-Maolegi
Jordan University of Science and Technology
{barraq_ghaleb@yahoo.com, mgm992002@yahoo.com}

## ABSTRACT
In an academic environment, student advising is considered a paramount activity for both advisors and student to improve the academic performance of students. In universities of large numbers of students, advising is a time-consuming activity that may take a considerable effort of advisors and university administration in guiding students to complete their registration successfully and efficiently. Current systems are traditional and depend greatly on the effort of the advisor to find the best selection of courses to improve students' performance. There is a need for a smart system that can advise a large number of students every semester. In this paper, we propose a smart system that uses association rule mining to help both students and advisors in selecting and prioritizing courses. The system helps students to improve their performance by suggesting courses that meet their current needs and at the same time improve their academic performance. The system uses association rule mining to find associations between courses that have been registered by students in many previous semesters. The system successfully generates a list of association rules that guide a particular student to select courses registered by similar students.

## 1. INTRODUCTION
Building smart software systems are emerging for medium to large scale systems where large amount of data needs processing. Data processing in many systems needs manual processing by supervisors and administrators of the system. For example, large universities need efficient systems to process the advising of a large number of students. The manual processing of students' schedules requires a large number of supervisors' efforts. The process of determining the courses, should students take next semester, should be finished in a few days. Such procedure is inefficient when dealing with large number of students. Moreover, advisors usually use a prescribed and general schedule for the students while not considering individual differences among students. Therefore, there is a need to have a smart system that can automate the decision making to help students in their registration process. The main focus of this research is to build a smart system that optimizes the registration process and reduces the effort of the university administration, advisors and students. The current advising system is a traditional and paper-based; besides, it consumes a considerable amount of time and effort from the department administration and advisors. Moreover, the process is as inaccurate as the available information to the advisor is not always sufficient. The course-load is a main concern for students and it needs to be considered carefully. Several factors may affect the course-load namely, the nature of the subject, student performance in prerequisite courses, student preferences, and the other courses taken by the student in the same term. These factors are not considered in the current advising system, therefore, the need for automated, and smart academic advising system is a must. The system will serve the following objectives:

− Projecting efficient use of resources in terms of instructors and classrooms before the registration period starts.
− Helping students register for classes according to their performances.
− Helping students with academic problems to improve their performances.
− Drawing conclusions and associations between courses.

To meet these objectives, we propose to use association rule mining to find and rank a list of suggested courses that a particular student can consider before registering. Students usually consult their advisors and preceding students before selecting courses for the next semester. The system ranks courses based on records and behavior of other students. One of the most frequently used data mining techniques in many applications is the association rule mining. It has been used as an underlying technology to improve the decision making process. In this study, we used one of the association rule mining methods namely; the Apriori algorithm as one method to build our advisory system in addition to the weight scheme.

In this work, the rest of the paper is organized as follows. In section 2, we discuss previous work on academic advising. In section 3, the association mining is described. Section 4 discusses the research methodology.

## 2. RELATED WORK
Building an advisory system for students is one of the applications that can employ the data mining methods. In this application, students should choose between dozens of courses every time they want to register for a course. Many choices in this situation are available which make it hard for a student to choose a specific course. Advisors are often overloaded with too many students and they usually have not enough time to help them. Moreover, students might not be satisfied with the preferences provided by advisors. Thus, it is necessary to build a recommendation system to help the student in making decision about what they should register. Although this system is needed and vital to the process of registration, literature reveals that it is difficult to build a global or a standard advising system that works for all universities and institutions because of differences of regulations and rules among different institutions [1][2][3].

Many researchers tried to build advising systems to facilitate the process of registration for students. We can distinguish two paradigms for building advising systems: a paradigm that relies solely on the advisor and the other paradigm relies on both the student and the advisor. The first one is not interactive where the student relies on the advisor to suggest him/her courses to register on and stick with these suggestions. An example of this kind is [4]. Whereas the second paradigm is more interactive where the student provides his/her preferences and the advisor builds his/her recommendations based on these preferences [5]. This type is more common and usually provides better outcomes compared to the first one. The authors in [6] developed a Web-based advising system for CS and CE undergraduate students where the system takes as input students' preferences and builds its recommendations upon these preferences. The students input their preferences and then get their recommendations from advisors via a web browser. This facilitates the process of advising and makes the process faster and more reliable. (This system is not an automated system and needs students to provide their preferences in order to build its recommendation which is a shortcoming that limits the power of this system). In another attempt to build a smart advising system the authors of [7] developed and designed a prototype for a rule-based expert advising system with an object oriented database. The authors built their system upon two categories; student preferences and academic rules. The system takes as input all students preferences and then validates these preferences against system rules. Although this system is an advancement towards building a smart system by integrating students' preferences with a process of validating these preferences (which refines those preferences), the authors failed to utilize their expert system to build an automated system that automatically generates recommendations based on noticed patterns among previously registered courses taken by previous students.

## 3. ASSOCIATION RULE MINING

Association rules are widely used in a variety of research areas such as inventory control, networks, risk management, market analysis, and telecommunication [8]. The association rule mining has the ability to detect interesting relationships between data items that happen frequently together. The association rule mining is a well-investigated method for discovering interesting correlations and relations among items in large datasets [9] [10]. The association rule mining is defined as: Let L = $\{i1, i2, ,in\}$ be a set of items. Let $D$ be a set of transactions, where each transaction $T$ is a set of items such that $T \subseteq L$. Each transaction is a unique identifier by a transaction (TID). We say that a transaction $T$ contain $X$, a set of some items in $L$, if $X \subseteq T$. An association rule is an implication of the form $X \rightarrow Y$, where $X \subseteq L, Y \subseteq L$, and $X \subseteq Y = \phi$. A rule $X \rightarrow Y$ is assessed using two measures are used to assess; support and confidence. The rule $X \rightarrow Y$ has support $S$ in the transaction set $D$ if $S\%$ of transactions in D contain $X \cup Y$. The rule $X \rightarrow Y$ holds in the transaction set $D$ with confidence $C$ if $C\%$ of transactions in $D$ that contain X also contain $Y$. The rules that have a support and confidence greater than the user-specified minimum support (denoted as min-sup) and the user minimum confidence (denoted as min-conf) are called interesting rules. A set of items is referred to as an itemset. A k-itemset contains k items and it is associated with each itemset a counter that holds the support of that itemset. Frequent itemset has a support greater than the minimum support. Candidate itemset is the one that is expected to be frequent. The rules generation is split up into two separate steps: (i) finding all frequent itemsets that have support above a predefined minimum support value; (ii) generating the association rules by using the frequent itemsets. Many algorithms have been proposed to solve the association rule mining problem. AIS algorithm is the first algorithm that has been proposed to mine association rules [9]. This algorithm generates only 1-item consequent association rules. For example, the rules of the form X, Y$\rightarrow$Z will be generated, but the rules of the form X$\rightarrow$Y, Z will not be generated. The primary disadvantages of the AIS algorithm are that it creates too many candidate itemsets and at the same time this algorithm requires too many passes over the whole database which reduces the computation performance of the algorithm. In this study, we use Apriori algorithm in order to generate the association rules because it is more efficient and simpler than other algorithms. To compute the support of itemsets, the Apriori algorithm uses the breadth first search algorithm. In this research, we propose to use an efficient Apriori association mining algorithm. To generate the interesting rules, the algorithm makes multiple passes over the database [11]. It uses the frequent itemsets in previous pass to generate candidate itemsets for the next pass. At pass $k$, it generates the candidate $k$-itemsets from frequent $k$-1-itemsets and then scans the database to find the support of each candidate $k$-itemset. Itemsets with support above the minimum support constitute the frequent $k$-itemsets set while other itemsets are discarded. The process continues until no more candidate itemsets could be generated. Generating candidate itemsets in the Apriori algorithm consists of two steps; join step and prune step. In join step, the algorithm joins different $k$-1-itemsets to generate $k$-itemset if they share the first $k$-2 items. In prune step, the candidate $k$-itemsets that are generated in the join step and have infrequent $k$-1-subset of $k$-1-itemsets, are removed. The prune step exploits the Apriori property that guarantees that all nonempty subsets of frequent itemset must also be frequent.

## 4. RESEARCH METHODOLOGY

In this section, we discuss how we implement the Smart Academic Advisory System (SAAS) using association rule mining. Association rule mining is considered as one of the most important data mining techniques. Its ultimate goal is to extract interesting correlations, frequent patterns, associations or casual structures among sets of items in the transactional databases or other data repositories.

### 4.1 Data Collection
We have selected the college Computer Information Technology at Jordan University of Science and Technology as a subject of the system. The system uses a real data from the registration pool. The dataset consists of 1530 student records of many semesters.

### 4.2 Data Preprocessing
Before using the Apriori algorithm to discover patterns among courses, there are many preprocessing steps that have to be conducted on the data to prepare it for the mining process. These pre-processing steps convert the student data to be suitable for mining. The pre-processing steps are as follows:

1. After the target student has entered his number or ID, the system will find the major in which the target student is specialized and find all the students in that major. For example, if the student major is computer science, then the system should look for all students whose major is computer science.

2. For each student in that major (target student major), the system will create one transaction for each semester that the student had been registered in. The transaction consists of the courses taken by the specified student given that the course score is equal to or greater than 50. For example, suppose that the two students with IDs 1 and 2 are the set of students who are in the same department of the target student and they take the courses depicted in Table 1. Then the system will construct the transactions as depicted in Table 2.

**Table 1: Students Registration**

| Student ID | The semester | Course name | Grade |
|---|---|---|---|
| 1 | First-Semester-2010-2011 | A | 50 |
| 1 | First-Semester-2010-2011 | B | 78 |
| 1 | First-Semester-2010-2011 | C | 40 |
| 1 | Second-Semester-2010-2011 | D | 67 |
| 1 | Second-Semester-2010-2011 | E | 30 |
| 1 | Second-Semester-2010-2011 | F | 50 |
| 1 | Summer-Semester-2010- | G | 40 |
| 1 | Summer-Semester-2010- | H | 90 |
| 2 | First-Semester-2010-2011 | A | 90 |
| 2 | First-Semester-2010-2011 | C | 78 |
| 2 | First-Semester-2010-2011 | D | 56 |
| 2 | First-Semester-2010-2011 | F | 84 |
| 2 | Second-Semester-2010-2011 | G | 76 |
| 2 | Second-Semester-2010-2011 | H | 54 |
| 2 | Second-Semester-2010-2011 | B | 94 |
| 2 | Summer-Semester-2010- | L | 87 |
| 2 | Summer-Semester-2010- | Z | 67 |

**Table 2: Student transactions**

| Transaction ID | The Transactions |
|---|---|
| 1 | A,B |
| 2 | D,F |
| 3 | H |
| 4 | A,C,D,F |
| 5 | G,H,B |
| 6 | L,Z |

### 4.3 Association Rules Generation

In this work, associative classification is constructed in two stages as suggested in [12]. First, the system identifies all associations that have a significant frequency of occurrences according to some thresholds. Second, classification rules are generated to build a classifier. Rules generation is conducted as shown in the pseudo-code in Figure 1. The students' registration data are then transformed into transactional by generating a transaction for a student that includes the student ID and a list of courses in a particular semester. For example, if a student is registered in four semesters then four transactions are generated. These transactions are then mined using the Apriori algorithm to generate the association rules. The rules that have not-registered courses on the left hand side are filtered out. The rest of the rules are then sorted by the confidence level.

```
Input: Student Data in Excel Sheet (ES),
ID of Student to be advised (ST), the
Confidence and the Support
Output: Suggested Courses
Step1: Foreach student (St) in ES
       ff major of (St) is equal to major
of (ST) then
         Foreach semester (St) have been
registered in ES
           Add a transaction consisting of
      the courses in this semester whose
      scores (>= 50) to the transactional
      database (TDB)
Step2: Apply Apriori algorithm on the
(TDB).
Step 3: Foreach generated rule of the form
A→B, C
       if the student have never
registered A then
         Delete the Rule A→B, C
       else if the student have already
registered C then
          Modify the rule A→B, C to be
A→B.
Step4: Rank the resulting rules according
to their confidence
Step5: Extract the right-hand side of all
      the rules, course ID. Put the
      extracted courses in an ordered
      list of suggested courses (SCs)
```

**Figure 1. Rule generation for smart academic advising**

### 5. RESULT ANALYSIS

After creating the transactional database (transactions), the data become ready to be mined using the Apriori algorithm and the system can generate the rules that the target user can use to get recommendations about the courses to register. The number of the rules that should be generated depend on the support and the confidence as provided by the system administer or the target student. Figure 2 shows a snapshot of the generated transactions after running the system on real data while Figure 3 depicts a snapshot of the generated rules after applying the Apriori algorithm on the transactions.

As the reader can conclude, some of the generated rules in Figure 3 are not valid. These the rules have a course in the right-hand side that has not been taken by the target user or the rules that have a course in the left-hand side that has been taken already by the target user. Thus, these rules should be filtered out. For example,

look at the following rule: 9900990, 1731020→1731011

The course (9900990) has not been taken by the target student. The condition of this rule is incorrect and as consequently the rule is incorrect and it should be removed from the set of the generated rules. Other types of the rules can be filtered out of the form:
1732111→1743450, 9900990

The course (1743450) is already has been taken by the target student. Thus, this rule should only be modified in order to be correct by removing the course (1743450) from the left-hand side of the rule. The modified rule will look as: 1732111→ 9900990. A snapshot of the final generated rules after the final processing and filtering are presented in Figure 4. The results in Figure 4 show the associations between courses and can help both the advisor and the student in drawing conclusions about related courses.

```
1741500,1743390,1744210
1742620,1743100,1744390
1742300,1743510,1762200
902331,1732111
1743400,1743450,1743830
1731020,1733180,1741500,1743400,1762200
902331,1741500,1743320,1752010
642000,901020,1743450
902331,1744420
1741500,1742280,1762300
1742520,1743830,1744290,1752010,1762200
1743020,1743400,1744210
821370,1731020,1733180,1743510,1744510,1763200
1742000,1743300,1743320,1744390,1744500,1744910
1731020,1741500,1743450,1744210,1744510
1734420,1742620,1743510
1742300,1743390,1744300,1744540,1752010
902331
1712310,1743320,1743830,1744390,1744910
662120,1733180,1744541,1744920
```

Figure 2. Snapshot of the generated transactions

```
900990-->1731020,811111,1731011------ 13%
900990-->801012------ 9%
900990-->1741000,1731020------ 16%
900990,811111-->1731020------ 56%
900990,811111-->1731020,1731011------ 53%
900990,811111-->1731020,1731011------ 53%
900990,811111-->1731011------ 72%
900990,811111,1731011-->1731020------ 74%
900990,1731011-->1731020------ 70%
900990,1731011-->1731020,811111------ 21%
900990,1731011-->1731020,821280------ 14%
900990,1731011-->1731020,821280------ 14%
900990,1731011-->821280------ 20%
900990,1731011-->1731020,811111------ 21%
900990,1731011-->811111------ 28%
900990,1731011,821280-->1731020------ 73%
900990,821280-->1731020------ 53%
900990,821280-->1731020,1731011------ 50%
900990,821280-->1731011------ 69%
900990,821280-->1731020,1731011------ 50%
900990,1731020-->811111,1731011------ 27%
900990,1731020-->1731011------ 93%
900990,1731020-->811111,1731011------ 27%
```

Figure 3. Snapshot of the generated rules

Finally, the system is used to extract a list of suggested courses. These courses are sorted according to the confidence of the rules. For example, Figure 5 depicts a snapshot of the final suggested courses sorted by their confidence. In Figure 5, the system recommends the student to select the courses that have the highest confidences. This list gives priorities to courses and helps the student in case too many courses are available during the registration period.

```
1741500-->811111,801012------15%
1741500-->811121,801012------9%
1741500-->901010,801012------23%
1741500-->901010,1731020------7%
1741500-->1752010------16%
1741500-->901020------7%
1741500-->1731121,801012,811111------10%
1741500-->801012,1731020------11%
1741500-->1731011,1731020------7%
1741500-->1731020------18%
1741500-->1742000------16%
1741500-->801012------48%
1741500-->811111------20%
1741500-->1731121,901010------18%
1742300-->1752010------28%
1742300-->1762300------26%
1742300-->1742280------21%
```

Figure 4. A snapshot of the final filtered rules

```
1742010----50%
801012----48%
1752010----43%
1731121----39%
1743320----38%
1733180----36%
901010----29%
1744500----28%
1762300----26%
902411----26%
1742280----25%
1742520----23%
1712310----23%
811111----20%
1731020----18%
1742000----17%
```

Figure 5. Final suggested courses

## 6. CONCLUSION

In this research, we proposed a new smart advising system to help undergraduate students during the registration period. Using a data mining technique, a smart academic advising system is implemented to help undergraduate students to register from a list of courses. The system uses a real data from the registration pool of Jordan University of Science and Technology. Although the registration process is not automated, the proposed system can reduce the amount of effort the advisors should spend advising a large number of students. In addition, students can select courses based on association among courses which is similar to what they are used to do in real life.

## 7. ACKNOWLEDGMENTS
This project is supported in part by an IBM corporation award and Jordan University of Science and Technology. (Grant# 271/2011)